\documentclass[twocolumn,showpacs,preprintnumbers,amsmath,amssymb]{revtex4}

\usepackage{graphicx}
\usepackage{bm}

\begin{document}

\title{Thermodynamic Evidence for Nanoscale Bose-Einstein Condensation  in $^4$He Confined in Nanoporous Media}

\author{Keiichi Yamamoto}
\author{Yoshiyuki Shibayama}%
\author{Keiya Shirahama}
\affiliation{%
Department of Physics, Keio University, Yokohama 223-8522, Japan
}%

\date{\today}

\begin{abstract}
We report the  measurements of the heat capacity of $^4$He confined in  nanoporous Gelsil glass that has nanopores of 2.5-nm  diameter at pressures up to 5.3~MPa.
The heat capacity has a broad peak at a temperature much higher than the superfluid transition temperature  obtained using the torsional oscillator technique.
The peak provides a definite thermodynamic evidence for the formation of localized Bose-Einstein condensates (LBECs) on nanometer length scales.
The temperature dependence of heat capacity is well described by the excitations of phonons and rotons, supporting the existence of LBEC.
\end{abstract}

\maketitle

Bose systems in periodic or random potentials offer a wide variety of novel phenomena and are a topic of great interest in condensed matter physics.
In ultracold atoms subjected to a periodic potential, quantum phase transition (QPT) between a superfluid phase and a Mott-insulator phase has been reported~\cite{Greiner2002}.
Introduction of disorder in an optical lattice induces a state without long range phase coherence, which is considered to be a Bose glass~\cite{Fallani2007}.
Thin superconducting films~\cite{Merchant2001} and  Josephson-junction arrays~\cite{Geerligs1989} have also been extensively studied as a disordered Bose system.

\begin{figure}[t]
\begin{center}
\includegraphics[%
  width=0.7\linewidth,
  keepaspectratio]{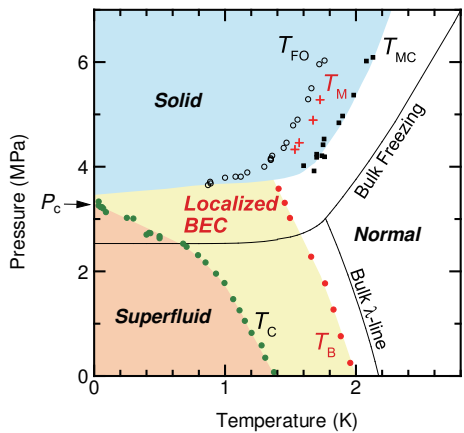}
\end{center}
\caption{The $P$-$T$ phase diagram of $^4$He in the 2.5-nm nanoporous glass determined in our present and previous studies~\cite{Yamamoto2004, Yamamoto2007}. 
The red circles and crosses indicate $T_\mathrm{B}$ and $T_\mathrm{M}$ in Fig. 1 and 2, respectively.
$T_\mathrm{c}$ indicate the superfluid transition temperatures obtained in the TO  study \cite{Yamamoto2004}, and 
$T_\mathrm{FO}$ and $T_\mathrm{MC}$ indicate the freezing onset and the completion of melting, respectively~\cite{Yamamoto2007}.
$P_\mathrm{c}$ indicates a critical pressure, 3.4 MPa.
The solid line shows the bulk $\lambda$-line and the bulk L-S boundary.
}
\label{PhaseDiagram}
\end{figure}

$^4$He in porous media is one of the ideal systems to study Bose-Einstein condensation (BEC) in an external potential.
By confining $^4$He in porous media, one can freely control the various experimental parameters such as dimension, topology, and disorder~\cite{Reppy, Shirahama1990, Wada2001, Chan1988}.
Recently,  we have discovered a novel behavior of $^4$He confined in a porous Gelsil glass by torsional oscillator (TO) studies~\cite{Yamamoto2004} and  pressure measurements (see Fig.~\ref{PhaseDiagram})~\cite{Yamamoto2007}, 
where the Gelsil glass is characterized by a three-dimensional (3D) random network of nanopores of 2.5-nm diameter.
The TO studies revealed that the superfluid transition temperature $T_\mathrm{c}$ is dramatically suppressed by pressurization.
At 3.4~MPa, $T_\mathrm{c}$ is suppressed to 0~K, which indicates the existence of a QPT at a critical pressure $P_\mathrm{c}$~=~3.4 MPa.

The pressure measurements revealed that the nonsuperfluid (NSF) state, which has very small entropy, exists between the superfluid and solid phase down to 0~K.
This observation implied that the low-entropy NSF phase in the confined $^4$He is possibly a localized state of BEC~\cite{Yamamoto2007}.
However, the entire  nature of the NSF state was not clarified by the pressure measurements.
Thermodynamic information is essential for understanding the novel NSF state.
In this Letter, we report  heat capacities  of  $^4$He confined in porous Gelsil glass, and demonstrate a definite evidence for the Bose-Einstein condensation on nanometer length scales in the nanopores of Gelsil glass above $T_\mathrm{c}$.

The Gelsil glass sample is a disk of 5.5-mm diameter and 2.3-mm height.
The samples are taken from the same batch as that employed in our previous studies~\cite{Yamamoto2004, Yamamoto2007}.
The sample cell is made of BeCu and one side of the cell wall acts as a diaphragm of an \textit{in-situ} Straty-Adams type capacitive pressure gauge.
The cell contains four Gelsil disks and has a space around the disk samples.
Thus, bulk $^4$He surrounds the glass samples in the measurements.
Temperature is  determined with Ge and RuO$_2$ bare chip resistors.
A  resistive strain gauge of 120 $\Omega$~\cite{Kyowa}  is glued to a face of the cell as a heater.
The cell is mounted on a massive isothermal Cu stage using a 4.4-cm long support tube made of a rolled Kapton sheet of 50~$\mu$m thickness,
where the stage is  located beneath the mixing chamber of a dilution refrigerator.
To suppress a superfluid counterflow of bulk $^4$He in a filling capillary, which causes the large heat leak from the cell, 
we insert a  Vycor superleak in the capillary~\cite{Yoon1997}.

We measure the heat capacities using the relaxation method or the adiabatic heat-pulse method, 
depending on the  external relaxation time of the cell to the thermal bath $\tau_1$ and the thermal relaxation time inside the sample $\tau_2$.
In the heat-capacity measurements of the empty cell,  $\tau_1/\tau_2$ are  6-7 above 0.45~K 
and  we employed the relaxation method~\cite{Shepherd1985}.
On the other hand, when we introduce $^4$He into the cell, $\tau_1/\tau_2$ exceeds 20 within the temperature and pressure range of measurements.
This condition is sufficient to employ the  adiabatic heat-pulse method.
We obtain the heat-capacity data by subtracting the empty-cell contribution, 
which is 3-31$\%$ of the heat capacity of $^4$He in the nanopores depending on temperature.

\begin{figure}
\begin{center}
\includegraphics[%
  width=0.75\linewidth,
  keepaspectratio]{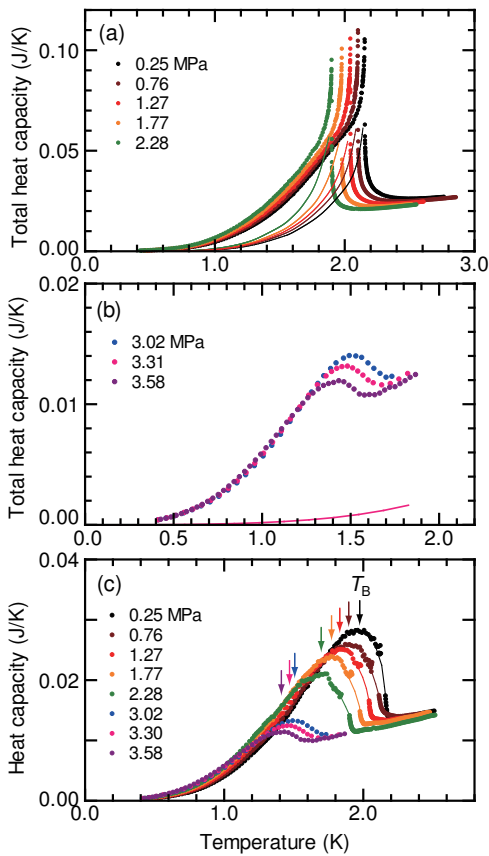}
\end{center}
\caption{Heat capacities at pressures where $^4$He in the nanopores is in the liquid state.
(a) Total heat capacity below 2.5~MPa.
The solid lines indicate the heat capacity of bulk $^4$He in the open volume of the sample cell.
(b) Total heat capacity at pressures at 3.02 $< P<$ 3.58~MPa where bulk $^4$He solidifies but $^4$He in the nanopores remains liquid.
The solid line is the heat capacity of bulk solid $^4$He in the open space at 3.31~MPa.
(c) The heat capacity after subtracting the contribution of bulk $^4$He  in the open space at 0.25~$<P<$~3.58~MPa.
The heat capacities show a broad peak at a temperature denoted as $T_\mathrm{B}$.
 The solid lines are guides to the eye. 
}
\label{T-C}
\end{figure}

We measure the heat capacities at pressures ranging from 0.25 to 5.28~MPa, and above 0.45~K.
At $P<$~2.5~MPa, the pressure in the nanopores is controlled by a room-temperature gas handling system.
At $P>$~2.5~MPa, we realized the heat capacity measurement along the isochore by using the blocked capillary method~\cite{Yamamoto2007}.
After the solidification of bulk $^4$He in the open space around the Gelsil disks in the sample cell, 
the $^4$He samples were carefully annealed by maintaining the cell within 50~mK of the bulk freezing temperature for 10 h 
to improve the crystalline quality of bulk solid $^4$He in the open space.
The pressure in the cell is measured by the low-temperature capacitive pressure gauge.

In Fig.~\ref{T-C}, we show the heat capacity data  at  various pressures; here, $^4$He in the nanopores is in the liquid state.
Figure \ref{T-C}(a) show the results below 2.5~MPa, where bulk $^4$He in the open space of the sample cell is in the liquid state.
The heat capacity shows a sharp peak around 2~K and a shoulder around 1.8~K.
The sharp peak originates from the lambda transition of bulk liquid $^4$He in the open space.
The solid lines indicate the heat capacities of bulk $^4$He obtained by using the existing specific heat data~\cite{Brooks1977, Lounasmaa1961}, where the quantity of bulk is estimated (e.g., 1.47~mmol at 0.25~MPa) using the sharp bulk-heat-capacity singularity at the lambda transition temperature $T_\lambda$ by employing  the method developed by  Zassenhaus and Reppy~\cite{Zassenhaus}.
The detailed procedure of the estimation is described elsewhere~\cite{YamamotoJLTP}.
At 0.25~$<P<$~2.28~MPa, the estimated quantity of the bulk liquid in the open space and the density of bulk liquid lead to a nearly constant volume of the open space of 38.4~mm$^3$, which is comparable to the value of 31~mm$^3$ estimated from the dimension of the cell and Gelsil disks. 

In Fig.~\ref{T-C}(b), we show the heat capacity  at pressures from 3.02 to 3.58~MPa, 
where $^4$He in the open space solidifies but $^4$He in the nanopores remains in the  liquid state~\cite{Yamamoto2007}.
The solid line is the heat capacity of bulk solid $^4$He at 3.31~MPa~\cite{Ahlers1970}, where the amount of bulk solid in the open space is evaluated  using  the estimated value of 38.4~mm$^3$ for the open volume and the density of bulk solid.
The contribution of bulk is smaller than that of liquid $^4$He in the nanopores.

The heat capacity data of liquid $^4$He in the nanopores obtained by the subtraction of the bulk contribution are shown in Fig.~\ref{T-C}(c).
All the data have a clear peak at a temperature denoted as $T_\mathrm{B}$.
As pressure increases, both $T_\mathrm{B}$ and the peak height decrease monotonically.
The heat capacity below 2.5~MPa resembles the data at higher pressures at which the heat capacity of bulk solid $^4$He  is negligibly small.
This fact justifies the subtraction of the heat-capacity contribution of bulk $^4$He at  $P<$~2.5~MPa.

\begin{figure}
\begin{center}
\includegraphics[%
  width=0.65\linewidth,
  keepaspectratio]{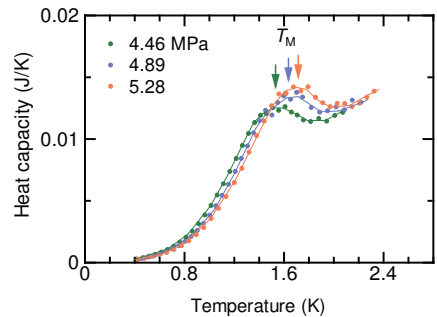}
\end{center}
\caption{Heat capacities of solid $^4$He in the nanopores of Gelsil after the subtraction of the small contribution of bulk solid $^4$He.
The broad heat capacity peak at a temperature denoted as $T_\mathrm{M}$ is caused by melting of $^4$He in the nanopores.
The solid lines are guides to the eye.
}
\label{T-C_subtraction}
\end{figure}

In Fig.~\ref{T-C_subtraction}, we show the heat capacities at $P>$~4.46~MPa, where $^4$He in the nanopores is solid~\cite{Yamamoto2007}. 
The heat capacity has a peak at  a temperature denoted as $T_\mathrm{M}$.
In contrast to the case of liquid $^4$He in the  nanopores, $T_\mathrm{M}$ and the peak height  \textit{increase} with  pressure.

The heat-capacity peak temperatures $T_\mathrm{B}$ and $T_\mathrm{M}$  are plotted on the $P$-$T$ phase diagram in Fig.~\ref{PhaseDiagram}.
The $T_\mathrm{B}$ line is shifted from the bulk lambda line by 0.2~K and is located  far above the superfluid $T_\mathrm{c}$ line.
The peak temperature $T_\mathrm{M}$  is located between the freezing onset and melting completion lines obtained from the pressure studies.

The heat capacity peak at $T_\mathrm{M}$ observed at pressures where $^4$He in the nanopores is in the solid state
 is attributed to the heat of melting of $^4$He in the nanopores.
In contrast to the first-order L-S transition in a  bulk system, the peak is rounded.
Peak rounding is a common feature of $^4$He confined in porous media~\cite{Brewer1990}.
Probably, the confinement of $^4$He  into the narrow pores and the pore-size distribution inherent in the porous glass cause the broadening of the L-S transition.

The heat capacity peak  of liquid $^4$He in the nanopores  at $T_\mathrm{B}$, which is shifted from $T_\mathrm{\lambda}$, may have been interpreted as a finite-size effect of the bulk superfluid transition, where the decreases in  the peak height and peak temperature are expected with decreasing the confining size~\cite{Gasparini1992}.
However, $T_\mathrm{B}$ is much higher than the superfluid transition temperature  $T_\mathrm{c}$ obtained by our previous TO study.
Therefore, the  peak  is not attributed to the superfluid transition  but some kind of short-range order.

We assert that  the broad heat-capacity peak indicates the formation of localized Bose-Einstein condensates (LBECs) on nanometer length scales.
LBEC was first suggested by Glyde and coworkers in  their neutron scattering experiment for $^4$He in Vycor and Gelsil at ambient pressure~\cite{Glyde2000, Plantevin2002}.
The LBEC originate from the spatial distribution of the critical temperature of BEC.
Both the confinement and  disorder produced by porous media may result in the novel LBEC state.
Below $T_\mathrm{B}$, many BECs start to form in favorable regions such as the largest pores,  in which $^4$He atoms can exchange their positions frequently. 
However, the exchange of atoms between BECs via unfavorable regions such as narrow pores is suppressed, because $^4$He atom has the hard core.
The suppression of atomic exchanges between the BECs results in the lack of global phase coherence in the whole system. 
Thus, the system does not exhibit superfluidity that is detected by macroscopic and dynamical measurements such as a TO.

In inelastic neutron scattering experiments, roton signals that are unique to BEC have been observed above the torsional oscillator $T_\mathrm{c}$~\cite{Glyde2000, Plantevin2002}.
Moreover, in $^4$He-filled Vycor, a broad heat capacity peak was observed above $T_\mathrm{c}$~\cite{Brewer1970, Zassenhaus}. 
These results and our Gelsil result are mutually consistent and are most probably attributed to the formation of LBEC in the nanopores.  

Theoretical studies have suggested that a random potential produced by a porous structure leads to the localized state of BEC.
Kobayashi and Tsubota studied the localization of BEC of  strongly correlated $^4$He confined in a 3D random environment.
They found that  superfluidity disappears above 4.2~MPa due to the pressure-induced localization of BEC~\cite{Kobayashi2005}.
A theory for a Bose gas with disorder at finite temperature suggests that the disorder leads to a temperature range where a locally condensed phase exists but superfluidity does not occur~\cite{Lopatin2002}.
Thus, these theories support the idea of LBEC.

\begin{figure}
\begin{center}
\includegraphics[%
  width=0.7\linewidth,
  keepaspectratio]{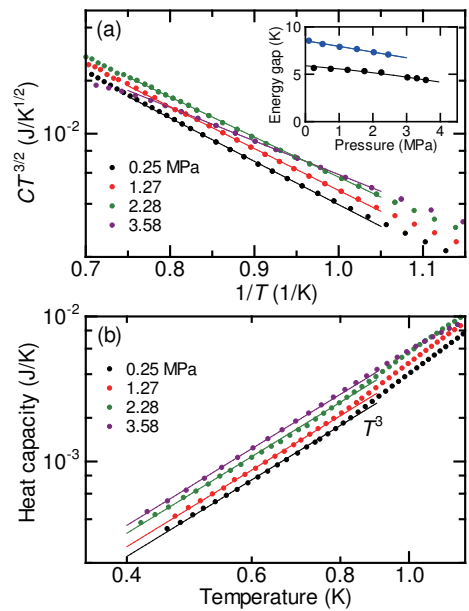}
\end{center}
\caption{(a) High-temperature heat capacities of liquid $^4$He confined in Gelsil.
The data are plotted as $CT^{3/2}$ vs $T^{-1}$.
Inset: Pressure dependence of roton energy gap, $\Delta$, of $^4$He in Gelsil (black).
The blue points show $\Delta$ of bulk $^4$He~\cite{Dietrich1972}.
(b) Log-log plots of heat capacity at low temperatures.
The solid lines are fitting of $T^3$ to the data between 0.4-0.7~K. 
}
\label{Phonon_Roton}
\end{figure}

No signature was observed in the heat capacity around $T_\mathrm{c}$ within the resolution of our measurements.
In $^4$He filled in Vycor, on the other hand, a very small heat-capacity peak was observed at $T_\mathrm{c}$, which is 0.02$\% $ of the size of the background heat capacity, in  high-resolution measurement~\cite{Zassenhaus}.
The resolution of our measurements is approximately 1$\%$ and might be insufficient to resolve the small peak at $T_\mathrm{c}$ if the peak exists.

The abovementioned picture of LBEC is also supported by the behavior of low temperature heat capacity below $T_\mathrm{B}$.
Above 1~K, the observed heat capacity is well explained by the excitation of the rotons.
The roton heat capacity is described by $C=AT^{-3/2} $exp(${-\Delta/k_{B}T}$), where  $A$ is constant and $\Delta$ is the roton energy gap.
In Fig.~\ref{Phonon_Roton}(a), we plot $CT^{3/2}$ versus $T^{-1}$.
The heat capacity data are well fitted as shown by the solid lines.
The pressure dependence of $\Delta$ are shown in the inset of Fig.~\ref{Phonon_Roton}(a).
As in the case of bulk $^4$He~\cite{Dietrich1972}, $\Delta$ decreases monotonically with increasing pressure.
$\Delta$ has a finite value 4.38~K at 3.58~MPa.
Importantly, the heat capacities at $P>$~2.5~MPa are well fitted by the roton heat capacity term.
Because the superfluid transition temperature $T_\mathrm{c}$ is well below 1~K, this observation proves that rotons are present well above $T_\mathrm{c}$.
This clearly shows the existence of LBEC in the nanopores above $T_\mathrm{c}$ and at $P>$~2.5~MPa.

At temperatures below 0.7~K, the roton contribution to the heat capacity vanishes.
We have found that in this temperature regime, phonons are the significant excitations.
In Fig.~\ref{Phonon_Roton}(b), we show the log-log plot of the heat capacities at $P<$~3.58~MPa.
The  heat capacities are well fitted by $T^3$ below 0.7~K.
This suggests that the heat capacity at low temperatures originates from 3D phonons.
According to Singh and Rokhsar~\cite{Singh1994}, 
the connectivity of the network of nanopores is important for excitation with wavelengths greater than the typical pore size $\lambda_\mathrm{pore}$.
Below a temperature $T_{x} \simeq hc/k_\mathrm{B}\lambda_\mathrm{pore}$, where $c$ is the  sound velocity in a liquid, 
phonons with wavelengths greater than $\lambda_\mathrm{pore}$ are thermally excited and the 3D behavior is expected.
If we use  the sound velocity in bulk $^4$He, $c=2.39 \times 10^4~\mathrm{cm/s}$, and $\lambda_\mathrm{pore}\sim$~1.5~nm which is the actual pore diameter obtained by assuming 1.5 inert layers of atomic $^4$He on the pore walls~\cite{Yamamoto2004}, we obtain  $T_x$~=~0.75~K.
This result is consistent with our observation that the  $T^3$ heat capacity appears below 0.7~K.
It is worth to mention that our most recent heat capacity measurements below 0.45~K show some deviation from the  $T^3$ to $T^2$ behavior.
This indicates that excitations other than phonons exist at lower temperatures.
Accurate measurements of heat capacity below 0.45~K will be essential in order to clarify the nature of the low-energy thermal excitations at lower temperatures.

At 0~K, the $P$-$T$ phase diagram implies that $^4$He confined in the nanopores undergoes the superfluid-LBEC-solid QPT at $P_\mathrm{c}$.
The physics of this QPT is not clarified in our heat capacity measurements because the lowest temperature of our measurements is 0.45~K.
The heat capacity study at lower temperatures near $P_\mathrm{c}$ is essential to reveal the entire nature of the QPT.

In conclusion, we have measured the  heat capacity  of $^4$He confined in a 2.5-nm nanoporous glass.
The heat capacity has a broad peak at temperature $T_\mathrm{B}$ that is much higher than  $T_\mathrm{c}$ determined using the torsional oscillator technique.
The heat capacities are well described by the excitations of phonons and rotons at all pressures. 
These observations prove the existence of nanoscale, localized Bose-Einstein condensates above $T_\mathrm{c}$.

This work is supported by the Grant-in-Aid for Scientific Research on Priority Areas "Physics of Super-clean Materials" from MEXT, Japan, 
and Grant-in-Aid for Scientific Research (A) from JSPS.
K.Y. acknowledges the support by Research Fellowships of the JSPS for Young Scientists.

\end{document}